\documentclass[sigconf, screen, authorversion]{acmart}
\usepackage[utf8]{inputenc}
\usepackage{csquotes}
\usepackage{amsmath}
\usepackage[english]{babel}
\usepackage{todonotes}
\usepackage[algoruled, noend]{algorithm2e}
\usepackage{array}
\usepackage{multirow}

\DeclareMathOperator*{\argmax}{arg\,max}

\AtBeginDocument{%
  \providecommand\BibTeX{{%
    \normalfont B\kern-0.5em{\scshape i\kern-0.25em b}\kern-0.8em\TeX}}}

\copyrightyear{2020}
\acmYear{2020}
\setcopyright{acmlicensed}
\acmConference[IUI '20]{25th International Conference on Intelligent User Interfaces}{March 17--20, 2020}{Cagliari, Italy}
\acmBooktitle{25th International Conference on Intelligent User Interfaces (IUI '20), March 17--20, 2020, Cagliari, Italy}
\acmPrice{15.00}
\acmDOI{10.1145/3377325.3377490}
\acmISBN{978-1-4503-7118-6/20/03}

\acmSubmissionID{1092}

\usepackage{enumitem}

\begin{document}

\title[AidMe]{User-in-the-loop Adaptive Intent Detection for Instructable Digital Assistant}

\author{Nicolas Lair}
\affiliation{%
  \institution{Inserm \& Cloud Temple}
  \country{France}
}
\email{nicolas.lair@inserm.fr}

\author{Clement Delgrange}
\affiliation{
  \institution{Cloud Temple}
  \country{France}
}

\author{David Mugisha}
\affiliation{%
  \institution{Cloud Temple}
  \city{}
  \country{France}
}
\author{Jean-Michel Dussoux}
\affiliation{%
  \institution{Cloud Temple}
  \city{}
  \country{France}
}
\author{Pierre-Yves Oudeyer}
\affiliation{%
  \institution{INRIA}
  \city{}
  \country{France}
}

\author{Peter Ford Dominey}
\affiliation{%
  \institution{Inserm}
  \city{}
  \country{France}
}

\renewcommand{\shortauthors}{Lair et al.}

\begin{abstract}
People are becoming increasingly comfortable using Digital Assistants (DAs) to interact with services or connected objects. However, for non-programming users, the available possibilities for customizing their DA are limited and do not include the possibility of teaching the assistant new tasks. To make the most of the potential of DAs, users should be able to customize assistants by instructing them through Natural Language (NL). To provide such functionalities, NL interpretation in traditional assistants should be improved: (1) The intent identification system should be able to recognize new forms of known intents, and to acquire new intents as they are expressed by the user. (2) In order to be adaptive to novel intents, the Natural Language Understanding module should be sample efficient, and should not rely on a pretrained model. Rather, the system should continuously collect the training data as it learns new intents from the user. In this work, we propose AidMe (Adaptive Intent Detection in Multi-Domain Environments), a user-in-the-loop adaptive intent detection framework that allows the assistant to adapt to its user by learning his intents as their interaction progresses. AidMe builds its repertoire of intents and collects data to train a model of semantic similarity evaluation that can discriminate between the learned intents and autonomously discover new forms of known intents. AidMe addresses two major issues – intent learning and user adaptation – for instructable digital assistants. We demonstrate the capabilities of AidMe as a standalone system by comparing it with a one-shot learning system and a pretrained NLU module through simulations of interactions with a user. We also show how AidMe can smoothly integrate to an existing instructable digital assistant.

\end{abstract}




\maketitle

\section{Introduction}
People are becoming increasingly accustomed to interacting with digital services, through vocal assistant or socialbots. Through cell phones, chatbot or connected speakers, it is possible to have a small conversation, to get answers to question or to take control of a connected device. These digital assistants range from a chatbot on an e-commerce website, that helps the user to get information on products, to a conversational agent that allows its user to have a conversation on topics such as sports, education or medicine. Natural language holds a huge potential for improving interactions of users with complex systems on the condition that it can provide useful features for the users while offering a seamless interaction.


From a user point of view, the experience with a digital assistant is as much influenced by the quality of the interaction as it is by the capabilities of the digital assistant. Much work remains to build a real conversational agent with fluent interaction skills. Two years ago Amazon launched the Alexa Prize. During the first Alexa Prize Competition~\cite{Ram2017}, the response error rate of the 15 developed conversational assistant ranged between $10$ and $30\%$. 

In most cases, digital assistants are specialized for certain tasks and their designers are allowed to maintain a certain quality of interaction on these specific tasks, entering in what Allen et al.~\cite{Allen2001} called \textit{set of contexts} systems. These virtual assistants are provided to the users with a full set of doable actions and their designers provide them with multiple ways to ask the assistant to perform these preregistered actions. Natural Language Understanding (NLU) and especially intent detection and slot filling can be achieved by tools like DialogFlow from Google, Wit.ai from Facebook. These assistants are static (i.e. non-evolving) and domain-specific by design.

The main limitations of these assistants is their inability to adapt to their users and to learn new tasks from their interaction with users. The few digital assistants~\cite{delgrange19, Liu+2016, allen_plow:_2007, IntharahHILC2017} that have been proposed to tackle these issues suffer from a weak comprehension capability: the assistant may experience difficulty in understanding user requests that are not expressed in the exact learned forms. As their task repertory grows and depends on the interaction they have with the user, they cannot rely on the traditional NLU tools mentioned above. Indeed such tools require the possible user intents to be hard-coded in the system from the beginning, and they rely on a large corpus of example sentences to train the models. 

In this work, we propose AidMe  – Adaptive Intent Detection in Multi-domain Environments – an NLU framework to allow an instructable assistant to leverage the performance of statistical machine learning models in the field of semantic similarity, without sacrificing their learning ability nor needing to engineer corpora of example sentences. AidMe is what we call a "Half-shot" learning system (between zero-shot and one-shot learning) able to integrate new intents from the users and autonomously discover new patterns of known intents as the interaction unfolds. AidMe trains an internal NLU model based on the collected sentences. In addition, AidMe is not domain-specific, and can be easily adapted as the NLU module of any instructable digital assistant. The purpose of AidMe is to demonstrate that it is possible to build customizable assistants without compromising on the quality of the interaction.
\smallskip

\textbf{Contributions.}
\begin{enumerate}
    \item We propose a NLU module that can be adapted to any Digital Assistant learning through interaction with a user, allowing the Digital Assistant to understand new intent from the user;
    \item AidMe is very sample efficient: it is a "half-shot" learning system;
    \item AidMe is by design a multi-domain NLU module.
    \item We propose a novel method to make use of semantic similarity to detect intent and patterns.
\end{enumerate}

\section{Related work}

Our work lies at the crossroads of three domains which we detail in the following section: intelligent assistant, natural language understanding and learning by interaction.

\subsection{Intelligent assistants}

As defined by Hauswald et al.~\cite{Hauswald2015}, an Intelligent Personal Assistant is \textit{"an application that uses inputs such as the user’s voice, vision (images), and contextual information to provide assistance by answering questions in natural language, making recommendations, and performing actions."} These assistants are used in many domains from medicine to education or e-commerce~\cite{GokselCanbek2016}.

While the performance in \textit{answering questions, making recommendations, and performing actions} is important, the user experience is paramount. One major limitation in the user experience is due to the quality of the interaction (e.g. the need for repetition when the assistant does not understand the request, etc.). Recent publications in the field of conversational agents attempt to address the difficulty of having a smooth conversation with an agent on diverse topics.  The introduction of Seq2Seq model~\cite{Sutskever2014} and its use on conversation task~\cite{Vinyals2015ICML} followed by~\cite{Li2016} have managed to provide conversational agent that can maintain consistent dialogue with different individuals.

When they are task-specific, these assistants can be completely tailored to meet a precisely designated purpose. Many tools such as DialogFlow from Google or Wit from Facebook allow almost anyone to quickly create a chatbot.

In the field of robotics, there has been significant research in advancing the skills of social robots  defined as \textit{embodied agents [...] able to recognize each other and engage in social interactions, they possess histories (perceive and interpret the world in terms of their own experience), and they explicitly communicate with and learn from each other.}~\cite{Dautenhahn1999}. Similarly as for Intelligent Personal Assistant, the quality and richness of the interaction between a robot and human is as important as its capabilities.~\cite{Fong2003}

\subsection{Natural Language Understanding}

\begin{table*}[!htbp]
  \caption{Example from ATIS dataset of slot filling: \\
  Show me the flights from Boston to New York today \\
  \textit{Intent}: flight\_search}
  \label{tab:atis_example}
  \begin{tabular}{c|cccccccccc}
    \toprule
    \textbf{Sentence} & Show&me&the&flights&from&Boston         &to&New             &York            &today\\
    \midrule
    \textbf{Label}        & O      &  O &O   &O        &O      &B.departure&O&B.arrival&I.arrival&B.date\\
  \bottomrule
\end{tabular}
\end{table*}

NLU consists in providing tools to automatically interpret and understand natural language. The first important task in NLU is \textit{intent detection}. It consists in being able to detect the intention underlying  a sentence or a paragraph. It is traditionally seen as a classification task with a huge number of classes that can prove to be extremely difficult depending on the number and nature of possible intents, the type of queries and the domain of application. For digital assistants, this is of critical importance as a poor intent detection system will prevent the assistant from providing relevant responses to the user.

In addition to intent detection, \textit{slot filling} or \textit{argument mapping} is another important task in NLU. Slot filling consists in semantically labelling each word of a sequence, helping the system to assign correct values to arguments. A classical example of slot filling with the sentence \textit{Show me the flights from Boston to New York today} is shown in Table \ref{tab:atis_example}. While these tasks were previously treated separately, recent research have shown that joint models capable of answering to both tasks performed better~\cite{Xu2013, Zhang2016, Firdaus2018a}.

Most common techniques are based on word embeddings and neural networks, especially recurrent neural networks such as LSTM and Attention based networks. Word embeddings are very convenient as they are learnt on wide corpora and can be tailored to a specific domain. Kim et al.~\cite{Kim2016}
showed how enhancing classical word embeddings such as Glove~\cite{pennington2014glove} to better represent similar and dissimilar words lead to very good performance on classical datasets with simple models.

A wide diversity of machine learning models have been used to create intent detection classifiers: ensemble model~\cite{Firdaus2018a, Figueroa2016}, convolutionnal networks~\cite{Xu2013}, recurrent network such as LSTM, bi-LSTM~\cite{Firdaus2018} or attention-based networks ~\cite{Liu+2016, Zhang2016}. All these methods perform extremely well, though it appears that attention-based methods outperform the others.

An essential assumption in these approaches is that the possible intents are known in advance. However in the case of a digital assistant that adapts to its user and learn from him, the different user's intents are discovered as he interacts with the assistant, not before. Thus we have access neither to the intents nor to a suitable train corpus, and training a classifier is not an option.

We should also mention the work of Allen~\cite{Allen2001} who argue that
using statistical methods to deal with the intent recognition process is not enough as they lack the ability to reason on the user inputs.

\subsection{Learning by interaction}\label{sec:learning-interaction}

The main limitation of domain-specific assistants is their inability to process requests that involve multiple domains or applications and to deal with actions which were not anticipated during system conception. To address this issue, Sun et al.~\cite{sun_intelligent_2016} developed an intelligent assistant framework that learns offline from the past interactions with the user to enable cross-domain usage of applications by suggesting relevant applications to the user's request. 

Another approach is to build instructable assistants that learn by interaction and can adapt to the user, such as HILC \cite{IntharahHILC2017} a user-in-the-loop system that learns procedures by user demonstrating and answering follow-up queries posed by the system. The system is quite promising, but does not address the complimentary problems of intent detection and argument mapping. 

PLOW, developed by Allen et al.~\cite{allen_plow:_2007}, converts natural language sentences into logical forms~\cite{hwang_episodic_1993} and thus needs a substantial engineering effort to be able to process sentences expressed with as much variability as in domain specific assistants. To avoid the limitations of systems based on grammatical rule engineering, systems such as LIA~\cite{azaria_instructable_2016} rely on data engineering. LIA parses natural language utterances by using a Combinatory Categorial Grammar (CCG) trained to associate natural language with a set of primitive functions. The system is able to derive more complex logical forms from user inputs, but as the authors pointed out, the system is hard to scale as it needs to discriminate between users' specific and common sense knowledge, which could be collected from different users to improve the parser.

\subsection{Semantic Similarity}

SemEval (International Workshop on Semantic Evaluation) offers competitions each year addressing themes related to the semantic evaluation of language. Between 2012 and 2017, one of these Semantic Evaluation Similarity (STS) competitions~\cite{DBLP:conf/semeval/2017} consists in developing a system capable of scoring the semantic similarity of a pair of sentences on a scale from 0 to 5, from completely dissimilar to perfectly similar. 

The techniques used are diverse and the results obtained are encouraging. Some apply neural network algorithms such as attention mechanisms~\cite{DBLP:conf/semeval/HendersonMSZ17} or convolutional networks~\cite{Shao2017}. Others compute variables from more traditional semantic and syntactic analysis tools such as alignment measures to feed supervised learning models~\cite{DBLP:conf/semeval/MaharjanBGTR17}. Teams~\cite{DBLP:conf/semeval/HassanABF17, DBLP:conf/semeval/WuHJGS17} also chose to rely only on an analysis of semantic tags from WordNet~\cite{Miller1993} and BabelNet~\cite{NavigliPonzetto:12aij}. And finally most of the teams chose to combine a few of these methods in an ensemble model.


\section{Modeling the approach}

\begin{figure}[!htbp]
  \centering
  \includegraphics[width=\linewidth]{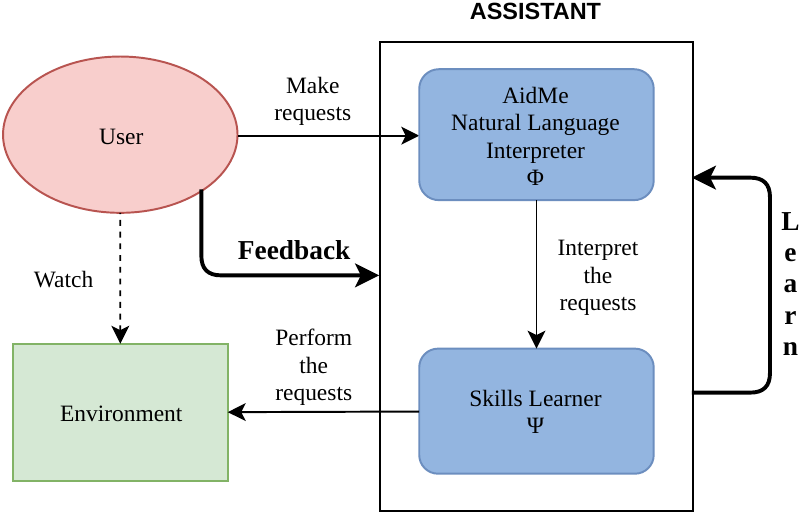}
  \caption{General architecture of an assistant learning by interaction: the assistant is divided into the Natural Language Interpreter and the Skill Learner}
  
  \Description[General architecture of an assistant learning by interaction]{General architecture of an assistant learning by interaction}
  \label{fig:model_assistant}
\end{figure}

In this section, we propose a general model of assistant that learns from interaction. As shown in Figure \ref{fig:model_assistant}, we distinguish between the Natural Language Interpreter and the Skills Learner to focus in the rest of the article on the Natural Language Interpreter.

\subsection{Model of a general assistant learning by interaction}

We focus on systems that can learn to match sentences to actions, through their interactions with a user, without prior knowledge of the domain nor the actions. We are then interested in systems that can fit in the following formalisation : 

\par To interact with the system, the user expresses its intent by using a sentence. Let $\mathcal{L}$ be the set of user's requests, $\mathcal{S} \subset \mathcal{L}$ the set of encountered sentences, $\mathcal{I}$ the set of learned intents, $\mathcal{P}$ the set of discovered patterns and $\mathcal{A}$ the set of actions doable by the assistant. Taking the sentence from Table \ref{tab:atis_example}, \textit{Getting the flight schedule} would be an intent, \textit{Show me the flights from Boston to New York today} would be a corresponding sentence, \textit{Show me the flights from DEP to ARR DATE} would be the corresponding pattern. We distinguish between an intent and a pattern as a unique intent can be expressed using many forms (and many patterns). However, there is an \textit{instantiation function} that links a sentence to its pattern: $\lambda: \mathcal{P} \times Args \rightarrow \mathcal{L}$, $Args$ being a dictionary of arguments used in the pattern to instantiate a sentence. 

\par We define two functions $\Phi : \mathcal{L} \rightarrow \mathcal{I} \times Args$, the natural language interpreter that guesses the intent lying behind a user's request and labels the word in the sentence, and $\Psi : \mathcal{I} \times Args \rightarrow \mathcal{A}$ that performs the requested actions depending on the intent and its arguments. Both $\Phi$ and $\Psi$ are learned during the interactions. We suppose that as the interaction between the system and the user unfolds, the user is requested to provide some guidance when the systems fails to fulfill the user's request. When the failure is especially due to the Natural Language Understanding module, the user's feedback should allow the system to compute the right intent and pattern.

This formalisation requires specific feedback from the user to be able to learn from  mistakes. Otherwise the system is left purposely general, no assumptions are made on $\Psi$ so that almost any system learning from interactions would be compatible. As illustrated by Figure \ref{fig:model_assistant}, our contribution aims at providing a general framework that can be used as the Natural Language Interpreter ($\Phi$) for digital assistants using its own Skill Learner and Environment.

\subsection{The Natural Language Understanding module}

AidMe is a Natural Language Understanding module that is able to learn to understand the user, meaning that depending on what is asked by the user, AidMe builds a repertoire of intents that the Skill Learner can learn to achieve. $\mathcal{S}$, $\mathcal{I}$ and $\mathcal{P}$ begin as empty sets and grow as the system interacts with the user. In this model, $\mathcal{S}$ can be considered as an \textit{interaction} memory, $\mathcal{I}$ and $\mathcal{P}$ as the \textit{learning memories}.

The obvious way to build such a system, would be to ask for the guidance of the user each time a request cannot be interpreted using the known patterns $\mathcal{P}$~\cite{delgrange19}. This would happen in the case where many patterns match the user's request or no pattern matches it. This is however what occurs in most of the interaction as the user rarely uses the same forms to express his intents. This method forces the user to always correct his assistant, rendering the system difficult to use. 


\section{Adaptive Intent and Pattern Detection}
\label{sec:aidme}

\begin{figure*}[!htbp]
  \centering
  \includegraphics[width=0.75\linewidth]{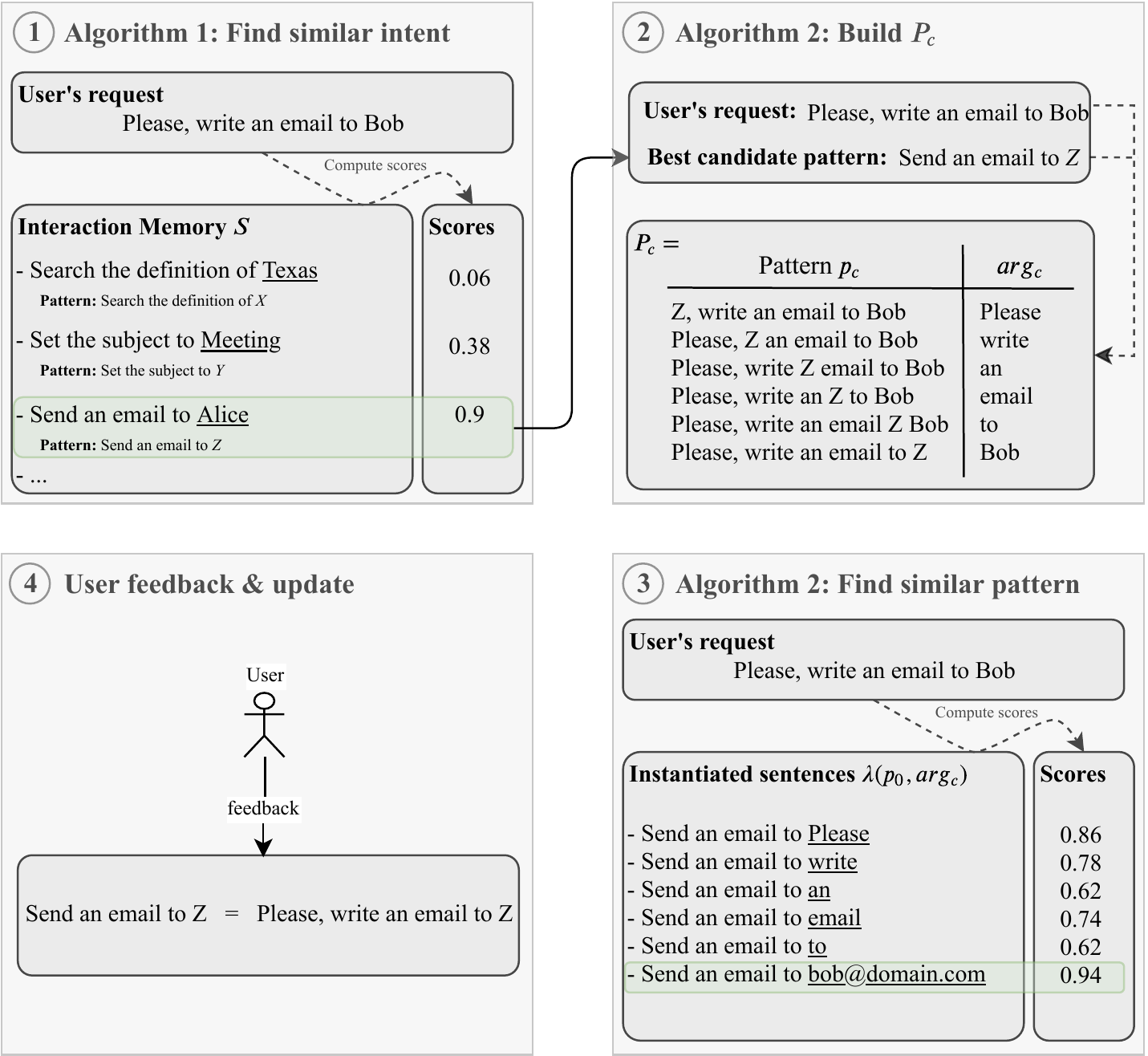}
  \caption{How AidMe analyzes the user's request to identify the intent, pattern and arguments in $4$ steps: (1) Identify the closest intent and pattern to a user request, (2) Build all the possible patterns from the user request (3) Identify the most meaningful pattern, (4) Update according to the user feedback }
  \label{fig:aidme_example}
  \Description{This figure describes }
\end{figure*}

In this section, we introduce AidMe and show how to take advantage of semantic similarity evaluation to perform semi-automatic intent and pattern identification with much less intervention of the user, and without sacrificing the learning ability of the assistant. Figure \ref{fig:aidme_example} shows the different steps performed by AidMe to analyze a user's request and identify its intent expressed through a new pattern.

\subsection{Similar intent identification}

\subsubsection{Problem description}

Intent identification is generally considered as a classification problem, each intent being one class. In the case of an adaptive system, one cannot know in advance the number of intents that the system will have to learn.   Additionally, multi-class classification with a high number of classes is difficult to address with traditional classifiers, and computationally expensive when using neural networks. To avoid these issues, we turn to the field of semantic similarity and reformulate the intent identification problem as trying to determine the semantic similarity between two sentences.

Intent recognition can indeed be seen as an indirect problem of semantic similarity evaluation: when an unseen sentence is addressed to the assistant, it is either referring to an already known intent or a to a completely new intent. The system has to discriminate between the two options. 

\subsubsection{Semantic Similarity based Intent Detection}

To do so, the system can search for the intent that would best describe the user's request. We present in Algorithm \ref{alg:intent_detection} our semantic similarity based intent detection process: we compare the user's request, $s \in \mathcal{L}$, to the known sentences and use the closest one to infer the intent underlying $s$. We hypothesize that two close requests have a similar intent. Given a model of evaluation of the semantic similarity (described in Section \ref{sec:sem_sim_model}) between two sentences $Sim: \mathcal{L} \times \mathcal{L} \rightarrow [0, 1]$, we look for $$s_0 = \argmax_{s' \in S} Sim(s, s') \text{ s.t. } Sim(s, s') \geq \epsilon $$ where $\epsilon$ is a confidence threshold above which the similarity is considered with high probability. The function $Sim$ can be interpreted as the inverse of a \textit{semantic distance} between two sentences in the sentence space $\mathcal{L}$, even if it does not have all the characteristic of a mathematical distance.
\begin{itemize}
    \item \textbf{If $s_0$ exists}, it is the closest sentence to $s$ that can be considered similar with a high enough probability and therefore $\Phi(s_0)$ is the best intent candidate and will be considered as such by the system. This means that the user's request refers to something the assistant has seen before.
    \item \textbf{If $s_0$ does not exist}, then no sentence seems similar enough to the unknown one and the system can consider that this new sentence refers to a new intent from the user. The assistant turns to \textit{learning mode} and can ask the user to explain his request.
\end{itemize}

\begin{algorithm}[!htbp]
\SetAlgoLined
\KwIn{User request $s$}
\KwOut{Identified intent $i \in I$, closest sentence $s_0$}
    $s_0 = \argmax_{s' \in \mathcal{S}} Sim(s, s')$\;
    $i =$ intent corresponding to $s_0$\;
    \If{$d(s, s_0) \geq \epsilon$}{\KwRet{$i, s_0$}}
    \Else{\KwRet{ $\emptyset, \emptyset $} (New intent detected!)}
 \caption{Semantic Similarity based Intent Detection}
 \label{alg:intent_detection}
\end{algorithm}

Although such a model would typically require data in order to be trained, here the data can be gathered through the interaction  with the user, and thus there is no engineering required to pre-specify the intents, as they are automatically discovered by the assistant. In addition, we will demonstrate that the sample efficiency is very good and thus that the performance of the method can rapidly become satisfactory. Finally the method is completely agnostic to the type of model and we can easily consider replacing a trained model by a model that does not need training.

\subsection{Similar pattern identification}

When the intent of a new request $s$ has been identified, the system still needs to identify the arguments before carrying out the request. The idea behind this step is to build the pattern $p$ corresponding to $s$ as the closest pattern to $p_0$ related to $s_0$. We define the contextual similarity $Sim_{|args}$ between two patterns $p_1$, $p_2$ with respect to a set of arguments $args$, by the semantic similarity between the two instantiated sentences: $$ Sim_{|args}(p_1, p_2) = Sim(\lambda(p_1, args),\lambda(p_2, args))$$
This definition is quite natural as it means that two patterns mean the same thing only if the corresponding sentences of these patterns mean the same thing.

\begin{algorithm}[!htbp]
\SetAlgoLined
\KwIn{User request $s$, closest sentence $s_0$, corresponding pattern $p_0$}
\KwOut{Identified pattern $p$}
Build $P_c = \{(p_c, args_c) \ | \ \lambda(p_c, args_c) = s\}$ \;
\ForEach{$(p_c, args_c) \in P_c$}{
$s_c = \lambda(p_0, args_c)$\;
$d(p_0, p_c) = Sim(s, s_c)$\;}
$new\_pattern = \argmax_{p_c \in P_c} d(p_0, p_c)$ \;
\KwRet{new\_pattern}
 \caption{Similar Pattern Identification}
 \label{alg:arg_detection}
\end{algorithm}

As described in Algorithm \ref{alg:arg_detection}, we can build the set $$P_c = \{(p_c, args_c) \ | \ \lambda(p_c, args_c) = s\}$$ of all possible pairs (patterns, arguments) that instantiate $s$ and have the same number of arguments as $p_0$. We finally use the model of evaluation of semantic similarity $Sim$ to measure the contextual similarity $Sim_{|args_c}$ between the pattern $p_0$ and the pattern in $P_c$ with respect to $args_c$. The closest candidate pattern to $p_0$ is assumed to be the pattern related to $s$. The idea behind this algorithm is that most of the instantiated candidate patterns will not even be correct sentences and thus have a lower similarity score than correct sentences and the correct pattern should get the highest similarity score.

\subsection{AidMe Algorithm}

\subsubsection{Interpreting the user's request}
Using the Algorithms \ref{alg:intent_detection} and \ref{alg:arg_detection}, we can build an algorithm that can identify most of the user's intents and allow the Skill Learner to carry them out. As shown in Algorithm \ref{alg:aidme}, the system first tries to interpret the user's request using the previous discovered patterns in $\mathcal{P}$. If it fails, the similar intent detection algorithm can determine if the request is close to known intent:
\begin{itemize}
    \item If not, the request is identified as referring to a new intent, the system has no chance to interpret it correctly and asks for the user's guidance, turning to \textit{learning mode}.
    \item If a close intent is identified, the system can infer the pattern corresponding to the request using the similar pattern identification pattern and then let the Skill Learner carry out the interpreted request. This case is described in Figure 2 with an example.
\end{itemize}

\begin{algorithm}[htbp]
\SetAlgoLined
\KwIn{User request $s$}
\KwOut{Identified intent $i$ and pattern $p$}
\eIf{$\Phi(s)$ is well-defined}
    {\KwRet{$\Phi(s)$}}
    {
    $i, s_0 =$ Algorithm 1$(s)$\;
    \eIf{$s_0$ exists}
        {
        $p_0$ : pattern corresponding à $s_0$\;
        $new\_pattern =$ Algorithm 2$(s, s_0, p_0)$\;
        \KwRet{$i, new\_pattern$}
        }
        {\KwRet{$\emptyset, \emptyset$} (New intent detected!)}
    }

 \caption{AidMe Algorithm}
 \label{alg:aidme}
\end{algorithm}

\subsubsection{Update of the system}
Once the sentence is interpreted, the systems can get feedback from the user. One can suppose that no feedback is a positive feedback, meaning the user is satisfied with the interpretation. If a feedback is given, we suppose that it includes the true intent and pattern. The system can then update $\mathcal{S}, \mathcal{I}$ and $\mathcal{P}$.

In order to improve its performance and to adapt to the user, AidMe periodically updates its semantic similarity model based on the collected sentences and intents. AidMe defines the semantic similarity using each intent as a class of similarity: 
$$Sim(s_1, s_2) = \begin{cases}
    1, & s_1, s_2 \text{ refer to the same intent}.\\
    0, & \text{otherwise}.
\end{cases}$$
Using this relation, AidMe can build a corpus of pairs of sentences and train its semantic similarity model.

We can note that the model does not need to be retrained each time a new intent is discovered to be able to recognize this new intent: two sentences can always be compared by the model even if one of the sentences refer to an intent that was not in the train set.


\section{Model of semantic similarity}

\label{sec:sem_sim_model}

In this section, we detail how we can compute a score of semantic similarity between two sentences. A model that evaluates the semantic similarity is a function $Sim: \mathcal{L} \times \mathcal{L} \rightarrow [0,M]$ where $M$ is an arbitrary integer. $Sim(s_1, s_2) = 0$ means totally dissimilar sentence, and $Sim(s_1, s_2) = M$ means perfectly similar sentences. We have developed such a model based on the works of the most successful teams in the Task 1 of SemEval 2017~\cite{DBLP:conf/semeval/2017}, and especially the ECNU team~\cite{DBLP:conf/semeval/TianZLW17}. The model used is an ensemble model whose sub-models can be separated into two groups: neural models using pre-trained word embedding vector representations such as Paragram~\cite{Wieting2015, Wieting2016} and decision tree models based on variables obtained from the analysis and comparison of each pair of sentences. The model is trained in a supervised manner on a set of sentences collected during the interaction of the system with the user. 

In this section we describe briefly each of the models and then evaluate the performance of our simplified model on the STS Benchmark dataset~\footnote{http://ixa2.si.ehu.es/stswiki/index.php/STSbenchmark}.

\subsection{Neural models}
\subsubsection{Description of the neural models}
We use multi-layer perceptrons taking a pair of sentences as input data. Each pair is converted into a vector representation as follows:
\begin{enumerate}
    \item Each word of each sentence is converted to its vector representation (word embedding);
    \item The embedding of a sentence is obtained by averaging the word embeddings;
    \item Finally, the embedding of a pair of sentences is computed by concatenating the Hadamard product (term to term product) and the $L1-distance$ between the sentence embeddings.
\end{enumerate}

The vector representations of the pairs then passes through two or three fully connected layers with rectified linear activation except for the last activation which is a softmax. The output layer of the network provides a distribution-like vector that encodes the similarity score. We use Paragram~\cite{Wieting2015, Wieting2016} as word embedding which is specifically engineered to deal with paraphrase and semantic similarity.

\subsubsection{Similarity score encoding}
In STS Benchmark used for SemEval competition, the semantic similarity is scored between $0$ and $5$. We consider two ways of encoding the semantic similarity score.
\begin{itemize}
    \item \textit{binary encoding}: The network outputs the probability of having a perfect similarity and the score is simply scaled according to the score range. Taking the STS Benchmark, a network output of $0.8$ corresponds a score of $4$. 
    \item \textit{$n$-class encoding}: Inspired by~\cite{Kiros2015, Tai2015}, each class encodes a degree of similarity and the network outputs a vector of size $n$ encoding the probability for each class. In that case the final score is the inner product between the model's output and the class vector $[0, 1, ..., n\text{-class}]$ (which can be seen as the score expectation of the computed score distribution). A score of $2.7$ would be encoded in the following vector $[0, 0, 0.3, 0.7, 0, 0]$, which can be interpreted as $p(similarity=2) = 0.3$ and $p(similarity=3) = 0.7$.
\end{itemize}

We implemented two neural models, one with a binary score encoding and another with a $n$-class score encoding. The particularity of these neural models is that they are based on the word meaning and do not take into account word order, syntax or grammar. We choose not to rely on model such as Long-Short Term Memory as their performance were considered disappointing on SemEval dataset~\cite{DBLP:conf/semeval/TianZLW17} and may be due to a wide variability between the training set and test set~\cite{Tai2015, Wieting2016}. In addition, they are much less sample efficient than perceptrons.

\subsection{Tree models}

 On the contrary, the tree-based models that we use take into account the syntax and grammar to compute different types of features. We use two types of ensemble tree models both based on binary decision trees: a random forest model and an xgboost model~\cite{Chen2016}. These models use features of two types derived from the pairs of sentences as inputs:

\subsubsection{Distance features} A first type of features that directly measures a distance or a similarity between the two sentences of the same pair:
    \begin{itemize}
        \item Measures of translation quality, like the BLEU score \cite{papineni-etal-2002-bleu};
        \item Several n-gram overlaps on characters and words;
        \item The Levenshtein distance (minimum number of insertions, deletions, or substitutions to transform one string into another).
    \end{itemize}

\subsubsection{Kernel features} A second type of feature is based on a vector representation of each sentence and a transformation using kernel functions. We use a Bag of Words (BoW) embedding on the vocabulary to transform each word into a vector. Each sentence is then the average of the word embedding weighted by its IDF (Inverse Document Frequency). The sentences are therefore represented in relatively large spaces (depending on the size of the vocabulary). To keep a balance between the different features and to avoid the curse of dimensionality when the vocabulary grows, each pair is transformed using the following kernel functions:
    \begin{itemize}
    \item linear kernel: cosine, manhattan distance, euclidean distance;
    \item statistic measures: Pearson correlation, Spearman correlation, Kendall tau;
    \item non-linear kernels: sigmoid, RBF (Radial Basis Function), laplacian.
\end{itemize}

The choice of a BoW embedding is justified by the natural sparsity of BoW which make the computation of the kernel extremely fast. With dense embedding, like pre-trained embedding as Paragram we use for neural networks, the computation of the kernel features is much more expensive. We experienced with embeddings like Paragram and Glove~\cite{pennington2014glove} and we found that the performance better than with BoW, but the little gain of performance could not compensate for the additional computation time and computation time is a key point for a digital assistant.


\subsection{The overall model}
\begin{figure}[htbp]
  \centering
  \includegraphics[width=\linewidth]{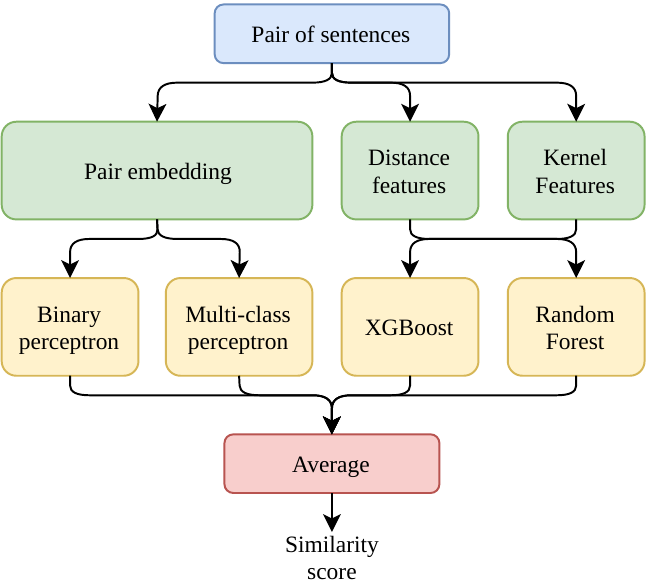}
  \caption{Ensemble Model of evaluation of semantic similarity: two neural networks using sentence embedding as inputs and two feature-based regression tree models}
  \Description[Model of evaluation of semantic similarity]{Model of evaluation of semantic similarity}
  \label{fig:ensemble_model}
\end{figure}

We use this four models (Binary perceptron (1), multiclass perceptron (2), xgboost (3), random forest (4) in an ensemble model by averaging the predictions of each model to get the model prediction. The model is illustrated in Figure \ref{fig:ensemble_model}.

\subsection{Model evaluation}

\paragraph{Comparison to SemEval baselines} Before applying the model in the context of AidMe, we have evaluated its performance on the STS Benchmark dataset~\footnote{http://ixa2.si.ehu.es/stswiki/index.php/STSbenchmark} used for SemEval competition. We evaluate on the test set a model that was trained on the train set. We show the results in Table \ref{tab:semeveal_comparison} and we report the some of results taken from Table 14 of the SemEval 2017 Task 1 Evaluation~\cite{Cer2017}. We report the results of three of the best teams and two baselines computed by the organizers: cosine similarity on sentence embedding obtained with InferSent~\cite{Conneau2017} or by averaging the word embeddings of a pre-trained word embedding (Paragram~\cite{Wieting2015}). We do not intend to compete with the best models but to ensure that our model is robust enough, stays close to the state of the art models while being computationally efficient. We see that our model is well above simple baselines and not too far from state of the art performance even though it could not beat InferSent which is very strong on this dataset. We show below that InferSent performs much worse on the intent detection task and we could not use it for AidMe.

\begin{table}[!htbp]
\captionsetup{justification=centering}
\caption{Comparison on the SemEval dataset between AidMe, baseline models and the best teams of the competition}
\label{tab:semeveal_comparison}
  \begin{tabular}{llc}
    \toprule
    \textbf{Model} & \textbf{Description} & \textbf{Pearson corr.}\\
    \midrule
    ECNU~\cite{DBLP:conf/semeval/TianZLW17}& Competitor&0.81\\
    BIT~\cite{DBLP:conf/semeval/WuHJGS17}& Competitor&0.81\\
    HCTI~\cite{DBLP:conf/semeval/TianZLW17}& Competitor&0.78\\
    InferSent & Organizers baseline & 0.76\\
    Paragram & Organizers baseline &0.50 \\
    \midrule
    AidMe model & & \textbf{0.75} \\
    \bottomrule
\end{tabular}
\end{table}


\section{Experiments}    

In this section, we present different simulations that prove the performance and adaptability of AidMe. The code of these simulations is available on \url{https://github.com/nicolas-lair/AidMe}

\subsection{UserGrammar}
\subsubsection{Simulation grammar}
\label{sec:User-Grammar} In order to evaluate AidMe, we build \textit{UserGrammar}, a grammar of 153 patterns, 50 intents and a dictionary of arguments. The different intents cover several domains such as mail, web search, concert booking, travel booking, calendar management and connected objects. We use this grammar to simulate user's requests on these domains. Table \ref{tab:grammar_ex} shows some examples of intent, patterns and instantiated sentences. The first two intents are identical, but the patterns are different and the sentences are differently instantiated. The nature of the arguments in the patterns are of different kinds: device, location, date, person... This is only used to generate meaningful sentences but AidMe has no idea of the nature of the arguments. The complete list of patterns and intent is available in the Supplementary Materials.

\begin{table*}[!htbp]
\captionsetup{justification=centering}
\caption{Examples of intent, patterns and sentences from UserGrammar}
\label{tab:grammar_ex}
  \begin{tabular}{lll}
    \toprule
    \textbf{Intent} & \textbf{Pattern} & \textbf{Example of sentence}\\
    \midrule
    Turn on a device somewhere & Can you turn the \_\_device on in \_\_loc &Can you turn the alarm on in Paris\\
    Turn on a device somewhere & Switch on the \_\_device in \_\_loc & Switch on the television in Paris\\
    Get next meeting with someone & When do I have a meeting with \_\_pers & When do I have a meeting with Bob\\
    Get cost for a flight on a day & Get me the cost for a ticket to \_\_loc on \_\_date& Get me the cost for a ticket to Berlin on Monday\\
    Get information &do you know what is a \_\_var& Do you know what is a spider\\
    \bottomrule
\end{tabular}
\end{table*}

\subsubsection{Model evaluation on UserGrammar}

To get an idea of how our model would perform on a task close to the user interaction simulation, we have evaluated our semantic similarity model on corpora generated with UserGrammar. We describe below the evaluation protocol. We generate a corpus from UserGrammar by generating:
\begin{itemize}
    \item A train set of 5000 pairs from 150 different sentences and 100 different patterns;
    \item A test set of 1000 pairs from 100 different sentences and 40 different patterns which were \textbf{not} used to generate the train set.
\end{itemize} 
We report the F1\_score, Precision and Recall of the models trained on the train set and evaluated on the test set. These metrics are more relevant than the Pearson correlation or a classic accuracy metric as the task is like a binary classification task with a highly unbalanced dataset (proportion of pairs with positive label between $2\%$ and $4\%$). We repeat the experiment 30 times and report the average performance of the model in Table \ref{tab:corpus_model_contribution} along with the standard deviation. It is important to note that \textbf{none} of the sentences nor the patterns of the test set were seen during training.

\begin{table}
\captionsetup{justification=centering}
\caption{Model evaluation on UserGrammar corpus}
\label{tab:corpus_model_contribution}
  \begin{tabular}{lcccc}
    \toprule
    \textbf{Model} & \multicolumn{2}{c}{\textbf{F1\_score}} & \textbf{Precision} & \textbf{Recall}\\
    {} & mean & std & {} & {} \\
    \midrule
    Binary Perceptron& 0.77 & 0.09 & 0.75 & \textbf{0.84}\\
    Multi-class Perceptron& 0.78 & 0.08 & 0.76 & \textbf{0.84}\\
    Random Forest & 0.82 & 0.08 & \textbf{0.88} & 0.77\\
    XGBoost & 0.80 & 0.09 & 0.85 & 0.77\\
    \midrule
    AidMe model & \textbf{0.85} & \textbf{0.05} & 0.87 & 0.82\\
    \bottomrule
\end{tabular}
\end{table}

We also report the average performance of each submodel to show the benefits of the ensemble model. We see that the feature-based models outperform the neural network on precision but are outperformed on recall. The ensemble model can make the most of it by outperforming each of the submodel in every metric. We note also that the standard deviation is almost half the standard deviation of the submodels, meaning more stability in the performance. A good balance between precision and recall is fundamental as good precision will prevent AidMe from wrong intent detection and a good recall will prevent AidMe from missing similar intents.

\subsection{Simulation of the user's interaction}
\label{sec:simulations}
\subsubsection{Simulation} Using the UserGrammar described above, we generate multiple sets of user's requests, one set is approximately 700 sentences and represent 700 interactions between a user and its assistant. At the beginning of each simulation, AidMe begins with no idea of what the intents or the patterns can be and the requests are presented to AidMe in a randomized order. AidMe has to determine the intent and the associated pattern, when it fails the simulator gives it the right intent and patterns so that AidMe can be updated. We illustrate this in Figure \ref{fig:aidme_example} on a example where AidMe analyses a sentence and integrates the user's feedback to update. In the simulation, we use $\epsilon=0.3$ for Algorithm \ref{alg:intent_detection}. This hyperparameter can be set depending if one wants to favor new or known intents detection accuracy.

In each simulation we can see three main phases: 
\begin{enumerate}
    \item \textbf{Intent learning phase}: Most of the user's requests refer to new intents or new patterns;
    \item \textbf{Pattern learning phase}: Most of the user's requests refer to known intents but new patterns;
    \item \textbf{Steady-state regime}: Most of the user's requests refer to known intents and patterns. 
\end{enumerate}

\subsubsection{Evaluation}

We evaluate two versions of AidMe using the two main following metrics the \textit{intent detection accuracy} and \textit{pattern detection accuracy} along with two baseline systems:
\begin{itemize}
    \item \textit{AidMe} is the algorithm as we describe it above
    \item \textit{AidMe\_M} is a variant of AidMe where a user request is always processed using Algorithms \ref{alg:intent_detection} and \ref{alg:arg_detection}. AidMe\_M does not try to match the request on a previously learned pattern.
    \item \textit{OneShotNLU}: a one-shot learning system that needs to see exactly each pattern once to be able to detect them using pattern matching. This system does not discriminate between intents and patterns and does not have a notion of similar patterns.
    \item \textit{DFLearner}: is a baseline system to compare AidMe with a system where the intent and pattern detection algorithms are replaced by another NLU system. We use DialogFlow\footnote{https://dialogflow.com/} API from Google, starting with a clear instance and adding intents and examples of sentences as the simulation unfolds. DFLearner cannot autonomously discover new patterns but every time new patterns and intents are encountered, they are added its knowledge base.
\end{itemize}

\subsection{Simulations' results}

We report the results of the following metrics in Table \ref{tab:simulation_result}: the accuracy in detecting new intents, knwon intents and the overall accuracy in intent detection. We report the same metrics for the pattern detection. As OneShotNLU does not make a difference between intents and patterns their intent detection accuracy and pattern detection accuracy are the same.

\begin{table}
\centering
\caption{Detection accuracy averaged over 10 simulations}
\label{tab:simulation_result}
\begin{tabular}{lccc|ccc} 
\toprule
\multicolumn{1}{r|}{\begin{tabular}[c]{@{}r@{}} \\ \end{tabular}} & \multicolumn{6}{c}{\textbf{Detection Accuracy ($\%$)} }                                                                                                                                                               \\ 
\cmidrule{2-7}
\multicolumn{1}{l|}{}                                             & \multicolumn{3}{c|}{\textbf{ Intents}}                                                                                        & \multicolumn{3}{c}{\textbf{Patterns }}                                                \\
\multicolumn{1}{c|}{\textbf{Model} }                              & New                                                    & Known                                                  & All         & New                                                    & Known         & All          \\ 
\midrule
OneShotNLU                                                        & 0                                                      & 80                                                     & 73          & 0                                                      & 100           & 73           \\
DFLearner                                                         & 0                                                      & 56                                                     & 52          & 22                                                     & 53            & 48           \\
AidMe\_M                                                           & 89                                                     & 91                                                     & 90          & 43                                                     & 70            & 61           \\ 
\midrule
\textbf{AidMe}    & \begin{tabular}[c]{@{}c@{}}\textbf{89}\\ \end{tabular} & \begin{tabular}[c]{@{}c@{}}\textbf{91}\\ \end{tabular} & \textbf{90} & \begin{tabular}[c]{@{}c@{}}\textbf{43}\\ \end{tabular} & \textbf{100}  & \textbf{82}  \\
\bottomrule
\end{tabular}
\end{table}

\subsubsection{Comparison with DialogFlow}
The performance of DFLearner are quite low as only half of the intent are correctly identified. This is because DialogFlow is not adapted to this kind of task:
\begin{itemize}
    \item DialogFlow needs to know the types of the arguments in the intent. As our system does not need the types of the arguments to detect new patterns and to be fair in the comparison we did not give the types of the argument to DialogFlow. A default argument type \texttt{@sys.any} exists but an intent cannot contain more than two non-typed argument. Besides, the types of arguments needs to be engineered in the system and it is not easy to think of a system that could learn them from the interaction. 
    \item DialogFlow is based on a combination of a rule-based and machine learning approaches. We got warning from DialogFlow that the number of example sentences were too small. This demonstrates why the indirect approach based on semantic similarity evaluation is relevant. With $n$ sentences, a direct classification model can build a corpus of size $n$ while our approach can build a corpus of size $n^2$.
\end{itemize} 
This experiment demonstrates that traditional NLU modules used to create dialogue assistants are not adapted to agents that can be instructed and customized by their user. 

\subsubsection{Comparison with OneShotNLU}
As OneShotNLU detects all the pattern when they have already been encoutered, the performance of OneShotNLU indicates the distribution between new and known patterns that appear multiple times. New patterns represents $30\%$ of the simulations and in $43\%$ of the cases, AidMe can detect this new pattern without needing the user's help. This means that the user intervention needed by a system using AidMe is almost halved compared to a system based on one shot learning.

However AidMe\_M has a lower pattern accuracy than OneShotNLU. This is due to the fact that the pattern detection algorithm reaches an accuracy of $70\%$ on the known patterns to be compared to the $100\%$ of OneShotNLU. This means the best combination is to use the pattern detection algorithm on new patterns and to perform pattern matching on known patterns.

\subsection{Integration to a digital assistant}

We integrate the AidMe algorithm to an existing virtual assistant described in~\cite{delgrange19} that learns by GUIs demonstrations and natural language instructions. The assistant is able to learn mapping of natural language orders to procedural knowledge in order to accomplish tasks on a digital environment composed of web services. The overall functioning of the system is as follows: the user gives an order, such as ``send an email to alice@domain.com'', and the system tries to find a corresponding procedure. If it does, the argument ``alice@domain.com'' is extracted and the procedure is executed, otherwise the system request from the user a GUI demonstration or an explanation of the procedure he wants to carry out.

A major limitation of this system is that the user has to remember the exact forms he used to teach its agent the new tasks. As an example, a request such as ``please, write an email to bob@domain.com'' is not recognized because the fixed pattern ``send an email to Z`` used to link natural language order to the procedure does not exactly match. The assistant then behaves as if the request referred to a completely new procedure and the user has to reformulate its request using the exact same form as during the learning. This can make the system hard to use. 

The AidMe algorithm, once integrated, allows the assistant to avoid around $40\%$ of these situations by proactively discovering new forms without behaving as if the request is entirely new. Using AidMe, the assistant can reformulate by itself the user's request in a known form and ask for confirmation to the user. When AidMe is wrong, in 90\% of the cases, it still has identified the intent correctly and can ask for reformulation. Below is an example of dialogue between the assistant and the user:

\begin{description}[leftmargin=!,labelwidth=\widthof{\bfseries Agent}]
    \item[User] Please, write an email to bob@domain.com 
	\item[Agent] Did you mean: send an email to bob@domain.com?	
	\item[User] yes 
\end{description}

As new procedures are learned and used, the system will gather the necessary data to improve the AidMe algorithm. This integration has improved the system's robustness to user language variability with a minimal set of pre-engineered language skills and still needs to be tested with more extended user experiments.

\section{Discussion}

\subsection{Pros and cons of a comparison-based method} 

\subsubsection{Pros: }A significant advantage of this method is that, such a model does not need to be trained each time a new intent is detected. Unlike traditional intent detection methods that are trained to recognize intents in a direct way and among a list of possible intents, our method is indirect as it seeks for a potential known intent without the assurance to find one. Two sentences can always be compared using the similarity model and the analysis of the similarity score is enough to evaluate if two sentences are close enough to be related to the same intent. On the contrary, a classification model can only predict classes on which it has been trained. When it discovers a new intent, a system based on a classification model needs therefore to be retrained. This can be an issue depending on the training time as the system would be unavailable for a while. But above all, the main challenge would be to maintain a relative balance in the dataset between classes, otherwise some class could be "ignored" by the model. When an intent is discovered for the first time, few data would be available to train on and the performance on the new class would need time to become satisfactory. Finally this method is more sample efficient than classification-based method as with the collection of $n$ sentences, the train dataset set is of size $n^2$.

\subsubsection{Cons: }The main disadvantage of this method is that the number of sentences to which a new request have to be compared grows rapidly. In addition, the pattern matching algorithm is challenged by very long sentences or sentences with many arguments. Detection of arguments is certainly a difficult task in NLU as the more arguments a sentence contains the bigger the number of possibilities grows. Our methods by comparing all the possible patterns is also vulnerable to the size of the sentences. As the number of words grows, the number of possible patterns explodes. We tracked the performance of AidMe in identifying the patterns depending on the number of arguments and report the results in Table \ref{tab:arg_pattern_discovery}. We note that it is extremely difficult for AidMe to discover new pattern that have more than 2 arguments.

\begin{table}[!htbp]
\centering
\caption{Pattern detection depending of the number of arguments}
\label{tab:arg_pattern_discovery}
\begin{tabular}{lccc} 
\toprule
                           & \multicolumn{3}{c}{\textbf{ Number of arguments}}  \\
\textbf{Pattern Detection ($\%$)} & 1  & 2  & 3                                        \\ 
\midrule
New Patterns  & 55 & 36 & 0                                        \\
Overall Accuracy   & 80 & 67 & 22                                       \\
\bottomrule
\end{tabular}
\end{table}

Upgrading AidMe to allow it to scale independently from the size of the collected sentences is part of the future work. One can think that once a certain number of sentences have been collected for an intent, it is not necessary to compare a new request to all of them but only to a small number. On the topic of making the similar pattern detection algorithm less vulnerable to the number of arguments and the size of the sentences, the idea that we should make use of the nature of the possible variables is very promising and would drastically reduce the combinatoriality of the approach.

\subsection{Sharing assistant knowledge across users}
\begin{figure}[!htbp]
  \centering
  \includegraphics[width=\linewidth]{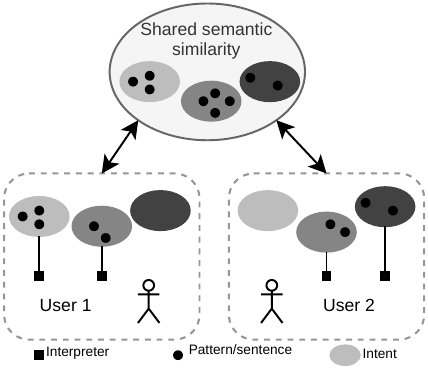}
  \caption{Shareable semantic similarity}
  \label{fig:shareable}
\end{figure}

As outlined in Section~\ref{sec:learning-interaction}, agents that learn by interaction gather data which are often related to a specific user. With a direct mapping between sentences and their interpretative function, it is hard , then, to share knowledge among users. It is common for a mapping to be not fully generalized and to hide details in the interpretative function. For example, a user may want to send emails with Application 1 while another user uses Application 2 but still say the same sentence ``send an email to X''. In our approach, as depicted in Figure~\ref{fig:shareable}, users may have different experiences with their agent but have the opportunity to share the linguistic similarity module to benefit from other user experiences. This model could be validated in future works.

\subsection{Enhancing the similar pattern identification algorithm}

The idea behind the pattern matching algorithm is to evaluate a semantic similarity between patterns by using the semantic similarity between instantiated sentences. In our implementation, we compare only the similarity between two patterns by instantiating one sentence for each pattern. We note in our simulations that this allows the system to infer around $40\%$ of the new patterns correctly. It would be interesting to evaluate how this can be improved by evaluating the similarity between patterns by averaging the contextual similarity over multiple contexts. Rather than comparing the similarity between two instantiated sentences, we could compare the average similarity between multiple instantiated sentences. We would then change the definition of the the contextual pattern similarity to:

$$Sim_{|ARGS}(p_1, p_2) = \frac{1}{|ARGS|} \sum_{args \in ARGS} Sim(\lambda(p_1, args),\lambda(p_2, args))$$

with $ARGS$ being a set of possible arguments associated with $p_1$ and $p_2$. We could expect the estimated similarity to be more accurate and the rate of discovered patterns higher. However, this needs to be balanced to the additional computation time that would require.

\section*{Future Work}

Along with multiple improvements to AidMe concerning its scaling abilities, it would be very instructive to test the system with real users by integrating it with different agents learning by demonstration. In addition to giving these assistants the possibility to autonomously infer the meaning of new sentences, AidMe can create novel types of interaction with the user. Able to detect new intents 90\% of the times, AidMe allows the assistant to be proactive in his interaction by indicating that it knows that the user is about to teach it a new task. In the same vein, when AidMe can fail to identify the patterns and but still indicates in more than 90\% of the times that it knows what the user wants but needs help to be able to perform it. This new type of interaction can make the interaction with digital assistant more human-like and attractive to the user.

\section*{Conclusion}

We developed AidMe as an NLU module for digital assistants that allows their users to teach them new tasks through interaction. To be usable, such a system needs to be extremely sample efficient and rapidly achieve good performance using only the data collected during the interactions. In this regard, AidMe achieves excellent sample efficiency. During the simulations we conducted, the intents were correctly detected $90\%$ of the times whereas AidMe has no more than 10 examples of sentences per intent at the end of the simulation. Additionally, AidMe beats one-shot learning systems by performing zero-shot learning on $43\%$ of the patterns in our simulation. AidMe is by design not a domain specific module, the knowledge that is pre-engineered in the system consists in the only word embeddings and this allows AidMe to adapt to the user in any domain. Finally AidMe can offer to the user a new sort of interaction where the assistants knows it needs the user to fulfill its request. The next step for AidMe is now to be confronted to real users by integrating it to real Digital Assistant.

\begin{acks}
Nicolas Lair is supported by ANRT/CIFRE contract No. 151575A20 from Cloud Temple.
\end{acks}
\bibliographystyle{ACM-Reference-Format}
\bibliography{biblio}

\newpage
\appendix

\end{document}